\documentclass[a4paper,12pt]{article}
\pdfoutput=1
\usepackage{graphicx, rotating}
\usepackage{hyperref}
\usepackage{slashed}
\usepackage{ifpdf}

\ifx\pdfoutput\undefined
\usepackage[dvips,bookmarks=false]{hyperref}	
\else
\usepackage{hyperref}	
\fi
\hypersetup{colorlinks,bookmarksopen,bookmarksnumbered,citecolor=verdes,
linkcolor=blus,pdfstartview=FitH,urlcolor=rossos}
\def\hhref#1{\href{http://arxiv.org/abs/#1}{#1}} 

\newcommand{\cm}{\,{\rm cm}}

\usepackage{multicol}
\usepackage{color}
\definecolor{rosso}{cmyk}{0,1,1,0.4}
\definecolor{rossos}{cmyk}{0,1,1,0.55}
\definecolor{rossoc}{cmyk}{0,1,1,0.2}
\definecolor{blu}{cmyk}{1,1,0,0.3}
\definecolor{blus}{cmyk}{1,1,0,0.6}
\definecolor{bluc}{cmyk}{1,1,0,0.1}
\definecolor{verde}{cmyk}{0.92,0,0.59,0.25}
\definecolor{verdec}{cmyk}{0.92,0,0.59,0.15}
\definecolor{verdes}{cmyk}{0.92,0,0.59,0.4}

\font\tenrsfs=rsfs10 at 12pt
\font\sevenrsfs=rsfs7
\font\fiversfs=rsfs5
\newfam\rsfsfam
\textfont\rsfsfam=\tenrsfs
\scriptfont\rsfsfam=\sevenrsfs
\scriptscriptfont\rsfsfam=\fiversfs
\def\mathscr#1{{\fam\rsfsfam\relax#1}}
\def\Lag{\mathscr{L}}

\newcommand{\fig}[1]{~\ref{fig:#1}}
\newcommand{\eq}[1]{~{\rm (\ref{eq:#1})}}

\newcommand{\GeV}{\,{\rm GeV}}

\def\circa#1{\,\raise.3ex\hbox{$#1$\kern-.75em\lower1ex\hbox{$\sim$}}\,}

\newcommand{\eqref}[1]{(\ref{#1})}

\newcommand{\beq}{\begin{equation}}
\newcommand{\eeq}{\end{equation}}

\def\circa#1{\,\raise.3ex\hbox{$#1$\kern-.75em\lower1ex\hbox{$\sim$}}\,}
\makeatletter

\def\art{\@ifnextchar[{\eart}{\oart}}
\def\eart[#1]#2#3#4#5#6{{\rm #2}, {#3 #4} {\rm (#6) #5} [arXiv:\-{\hhref{#1}}]}
\def\hepart[#1]#2{{\rm #2, arXiv:\-\hhref{#1}}}
\newcommand{\oart}[5]{{\rm #1}, {#2 #3} {\rm (#5) #4}}

\newcounter{alphaequation}[equation]
\def\thealphaequation{\theequation\hbox to
0.6em{\hfil\alph{alphaequation}\hfil}}
\def\eqnsystem#1{
\def\@eqnnum{{\rm (\thealphaequation)}}
\def\@@eqncr{\let\@tempa\relax \ifcase\@eqcnt \def\@tempa{& & &} \or
\def\@tempa{& &}\or \def\@tempa{&}\fi\@tempa
\if@eqnsw\@eqnnum\refstepcounter{alphaequation}\fi
\global\@eqnswtrue\global\@eqcnt=0\cr}
\refstepcounter{equation} \let\@currentlabel\theequation \def\@tempb{#1}
\ifx\@tempb\empty\else\label{#1}\fi
\refstepcounter{alphaequation}
\let\@currentlabel\thealphaequation
\global\@eqnswtrue\global\@eqcnt=0 \tabskip\@centering\let\\=\@eqncr
$$\halign to \displaywidth\bgroup \@eqnsel\hskip\@centering
$\displaystyle\tabskip\z@{##}$&\global\@eqcnt\@ne
\hskip2\arraycolsep\hfil${##}$\hfil& \global\@eqcnt\tw@\hskip2\arraycolsep
$\displaystyle\tabskip\z@{##}$\hfil
\tabskip\@centering&\llap{##}\tabskip\z@\cr}
\def\endeqnsystem{\@@eqncr\egroup$$\global\@ignoretrue} \makeatother

\begin{document}

\begin{center}
IFUP-TH/2009-27\hfill

\bigskip\bigskip\bigskip

{\huge\bf\color{magenta}
CDMS stands for\\[5mm]
Constrained  Dark Matter Singlet}\\

\medskip
\bigskip\color{black}\vspace{0.6cm}
{\bf 
\large Marco Farina$^a$, 
 {\bf Duccio Pappadopulo}$^b$  {\rm and} \bf Alessandro Strumia$^{cd}$}
\\[7mm]
{\it $^a$ Scuola Normale Superiore and INFN, Piazza dei Cavalieri 7, Pisa, Italy} \\
{\it $^b$ ITPP, EPFL, CH-1015, Lausanne, Switzerland} \\
{\it $^c$ Dipartimento di Fisica dell'Universit{\`a} di Pisa and INFN, Italia} \\
{\it $^d$ CERN, PH-TH, CH-1211, Gen\`eve 23, Suisse}

\bigskip\bigskip\bigskip\bigskip

{
\centerline{\large\bf Abstract}
\begin{quote}
Motivated by the two candidate Dark Matter events observed by the CDMS experiment, 
we consider a Constrained Dark Matter Singlet (CDMS) model
that, with no free parameters, predicts the DM mass and the DM direct cross section to be in the range weakly favored by CDMS.
\end{quote}}

\end{center}


\section{Introduction}
\label{intro}

The Cryogenic Dark Matter Search (CDMS) experiment reported 2 events possibly due to DM scattering,
with an expected background of about 0.8 events~\cite{CDMS}.
The statistical significance of the hint is so low, about $1.5\sigma$,
that calling it excess would be an excess.  
It is further reduced by the bound of the {\sc Xenon} experiment~\cite{Xenon}, that also observed events, interpreted as background.
However discoveries are fist seen as hints, and the CDMS excess attracted theoretical interest~\cite{CDMSth,DMS}.

The event rate as well as the average energy of the two CDMS events give an indication
on the DM cross section and on the DM mass:
DM lighter than the nuclear mass gives less energetic scattering events.
We determine the `favored' region by performing an event-by-event fit, assuming a dominant spin-independent
elastic DM/nucleon cross section $\sigma_{\rm SI}$,
the `standard'  local DM density $\rho_\odot = 0.3\GeV/\cm^3$
(one should take into account that only the combination $\rho_\odot\sigma_{\rm SI}$ is determined
and that values closer to $0.4\GeV/\cm^3$ might actually be favored by Milky Way rotation curves~\cite{UllioCatena})
and for the DM velocity distribution
(in the galactic frame, a Maxwellian 
${dN}/{d\vec v} \propto e^{-v^2/(220\,{\rm km/s})^2} \Theta(v-v_E)$
cutted at the escape velocity $v_E= 500$ km/s), 
and taking into account the energy-dependent CDMS signal efficiency
(the correct likelihood for this purpose was described in~\cite{Lik}).

\begin{figure}
\begin{center}
$$\includegraphics[width=0.65\textwidth]{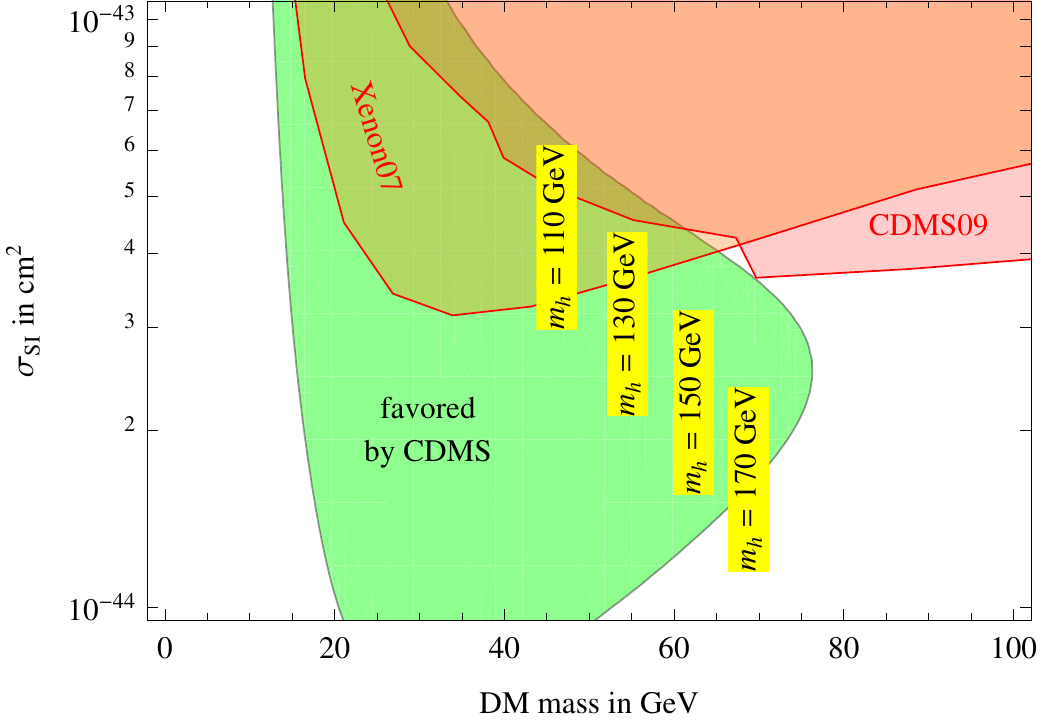}$$
\caption{\em Predictions of the CDMS model for spin-independent elastic scattering for $M$ and $\sigma_{\rm SI}$ for a few values
of the higgs mass $m_h$
compared to the region `favored' at  $78\%$ CL ($\Delta\chi^2<3$ for 2 dof) by the CDMS experiment (green region),
and to the regions disfavored at $90\%$ CL by CDMS and {\sc Xenon} (upper red shaded regions).
We assumed $\rho_\odot = 0.3\GeV/\cm^3$.
\label{fig:2}}
\end{center}
\end{figure}

Fig.\fig{2} shows the regions `favored' by the two CDMS events at $78\%$ confidence level (i.e.\ $\chi^2 - \chi^2_{\rm min}<3$ for
2 dof).
Our results agree with those in~\cite{Zupan}.
Of course, any significantly higher confidence level would favor the whole parameter space
not disfavored by CDMS and {\sc Xenon} data~\cite{Xenon}.
The two jumps in the bound reported by CDMS are due to the two events.
We see that the CDMS events suggest $\sigma_{\rm SI} \approx \hbox{few}\times 10^{-44}\cm^2$
and a DM mass $M$ around 40 $\div$ 80 GeV.

\bigskip

 Perhaps, proposing DM models that predict $M$ and $\sigma_{\rm SI}$ to be 
in the CDMS range, despite the weakness of its statistical significance,
could maybe have some possible interest.
At least, this is more interesting than proposing models where $M$ and $\sigma_{\rm SI}$ can lie in the CDMS
range, but suffer orders of magnitude uncertainties.

\section{The model}
The CDMS value of $\sigma_{\rm SI}$ is characteristic of higgs-mediated
DM/nucleon scattering.
Thereby we consider a DM model obtained adding to the Standard Model a Dark Matter real singlet scalar field $S$
coupled to the Higgs doublet $H$ as described by the following Lagrangian invariant under $S\to -S$:
\beq \Lag =\Lag_{\rm SM } + \frac{(\partial_\mu S)^2}{2} -\lambda S^2 |H|^2\ .\label{eq:L}\eeq
This is the well known scalar singlet model~\cite{McDonald, Burgess, Barger, DMS}, 
here with the additional constraint of setting to zero the mass term $m^2 S^2/2$
(and omitting the quartic coupling $S^4$, which is phenomenologically irrelevant).
Inserting $H=(0,V+h)/\sqrt{2}$ in eq.\eq{L}, the higgs vev $V\approx 246\GeV$ gives a DM mass $M=\sqrt{\lambda} V$.

Phenomenologically, $m$ cannot be much larger than the Higgs mass $m_h$.
Theoretically, $m$ (as well as $m_h$)
has no reason to be much smaller than the Planck scale; both receive quadratically divergent quantum corrections
so that their smallness is technically unnatural, giving rise to hierarchy problems.
As well known, the smallness of $m_h$ could be related to the scale of supersymmetry breaking,
with $m_h=0$ in the supersymmetric limit.
Maybe $m$ is similarly forbidden by some symmetry which does not need to be broken, such that $m=0$.
Alternatively, if $m$ and $m_h$ are small due to independent reasons, 
it is unlikely that they are comparable; one therefore expects that $m\ll m_h$ is `more likely'.
%

\smallskip

Whatever is its motivation, the model is phenomenologically interesting: it
has only one parameter, $\lambda$, fixed by assuming that thermal freeze-out reproduces the
observed cosmological DM abundance.
Thereby both $M$ and $\sigma_{\rm SI}$ are predicted, as we now compute.
As both predictions lie within the region `favored' by the CDMS experiment, we
name this model Constrained Dark Matter Singlet (CDMS).

\begin{figure}
\begin{center}
$$\includegraphics[width=0.45\textwidth]{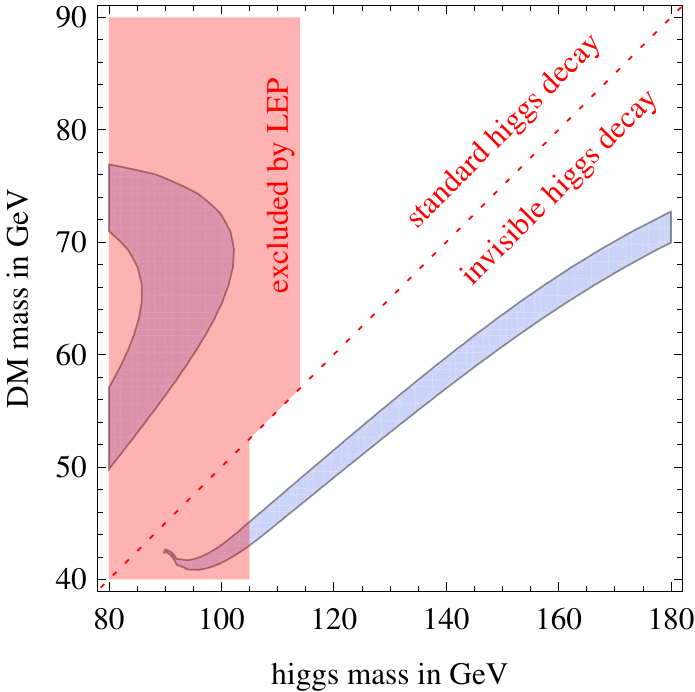}\qquad
\includegraphics[width=0.46\textwidth]{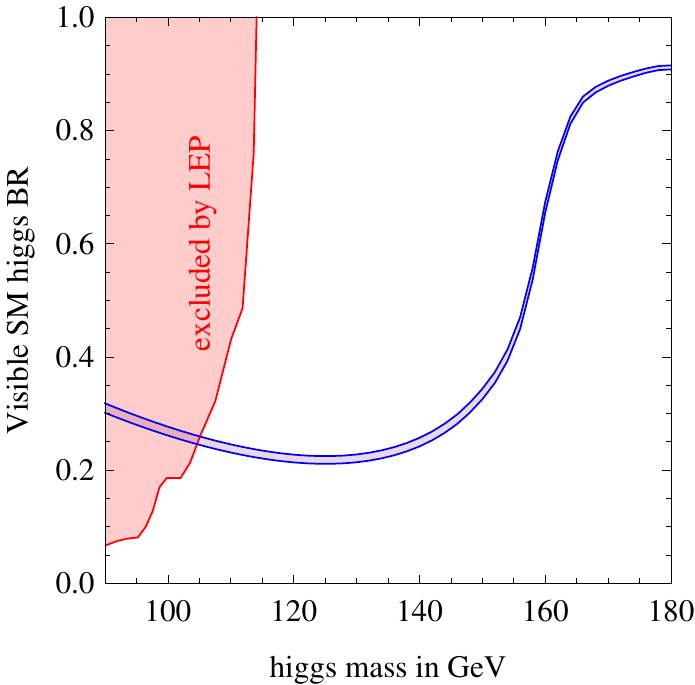}$$
\caption{\em Left: regions in the ($m_h,M$) plane compatible with the cosmological DM abundance.
Right: reduction in the SM higgs branching ratio due to the extra invisible decay $h\to SS$.
\label{fig:1}}
\label{default}
\end{center}
\end{figure}

\medskip

According to the standard approximation for cosmological thermal freeze-out, the DM relic abundance 
$\Omega_{\rm DM} =\rho_{\rm DM}/\rho_{\rm cr}$
is given by:
\beq\label{eq:OmegaDM}
\Omega_{\rm DM}h^2=\frac{1.1\,z_f}{\sqrt{g_{\rm SM}}M_{\rm Pl} \sigma v\,{\rm eV}},\qquad
z_f \approx \ln \frac{0.038 M_{\rm Pl}M\sigma v}{\sqrt{g_{\rm SM}} z_f}\approx \frac{1}{23}
\eeq
where $g_{\rm SM}\sim 80$ is the effective number of relativistic degrees of freedom at the freeze-out temperature  $T_f = Mz_f\sim 10\GeV$. 
The non-relativistic $s$-wave
$SS$ annihilation cross section into SM particles is given by~\cite{McDonald, Burgess, Barger, DMS}
\beq
\sigma v = \frac{8\lambda^2 V^2}{[4M^2 - m_h^2]^2 + m_h^2 [ \Gamma(m_h)+\Gamma_S]^2} \frac{\Gamma(2M)}{2M}\eeq
where $\Gamma(m)$ is the decay width of a Higgs boson with mass $m$ into SM particles, and
$\Gamma_S = \lambda^2 V^2\,{\rm Re}\sqrt{1-4M^2/m_h^2}/8\pi m_h$ is the higgs decay width into $SS$.



Fig.\fig{1} shows the values of $M$ compatible with the measured cosmological DM density,
$\Omega_{\rm DM} h^2 =0.110 \pm 0.005$~\cite{cosmoDM}.
We plot the $3\sigma$ band, assuming a $5\%$ uncertainty in the theoretical prediction of eq.\eq{OmegaDM},
that accounts for the $p$-wave contribution to $\sigma v$, suppressed by a ${\cal O}(z_f)$ factor.

We see that multiple solutions are possible: 
\begin{itemize}
\item[i)] with $m_h < 2M$, such that the higgs is standard, $\Gamma_S=0$. 
This solution exists only for $m_h\circa{<}115\GeV$,
and seems excluded by LEP searches, possibly unless $m_h$ is just around the LEP bound, $m_h > 114\GeV$~\cite{EWPT}.

\item[ii)] with $m_h>2M$, such that $h\to 2S$ decays are kinematically allowed.
Fig.\fig{1}b compares the prediction for the visible higgs branching ratio
with the LEP bounds: this solution is allowed for $m_h \circa{>}105\GeV$~\cite{hinv}.
The predicted coupling, $\lambda\sim 0.05$, is perturbative.
\end{itemize}
Furthermore, within the SM as well as within the CDMS model,
precision data favor $m_h = (87\pm 35)\GeV$~\cite{EWPT}, and, 
together with TeVatron higgs searches, suggest $m_h\circa{<}160\GeV$ at $90\%$ confidence level~\cite{EWPT}.

\medskip

We now compute the spin-independent DM/nucleon elastic scattering mediated by tree-level higgs exchange
(see also~\cite{McDonald, Burgess, Barger, DMS}).
The interactions $\lambda V S^2 h$ and $m_q h\bar{q}q/V$
give the effective operator
$\lambda S^2 m_q \bar q q/m_h^2$.
Its nucleon matrix element is
\beq \langle{N} | \sum_q m_q \bar{q}q | N\rangle\equiv  f m_N [\bar N N] \eeq
where the sum runs over $u,d,s,c,b,t$ and,
according to recent analyses, $f= 0.56\pm 0.11$~\cite{Ellis} or $f= 0.30\pm0.01$~\cite{lattice} using lattice results.
The prediction for the conventional spin-independent DM/nucleon cross section is:
\beq \sigma_{\rm SI}
= \frac{\lambda^2 m_N^4 f^2}{\pi M^2 m_h^4}\ .\eeq


\medskip

Fig.\fig{2} shows the numerical prediction for both $M$ and $\sigma_{\rm SI}$ as function of the Higgs mass $m_h$.
The central values assumes $f=1/3$ (as in~\cite{MDM}) and
the size of the bands indicates the  uncertainty in the nuclear matrix element.
The disfavored solution with $m_h\approx 115\GeV$ would give a larger $\sigma_{\rm SI}\approx 10^{-43}\cm^2$,
a value disfavored also by CDMS and {\sc Xenon} direct searches. Indirect DM signals are compatible with existing bounds.

\medskip

A related model is obtained assuming that the scalar singlet is complex under some dark U(1),
with Lagrangian $ \Lag =\Lag_{\rm SM } + |\partial_\mu S|^2 -2\lambda |S|^2 |H|^2$.
As this model is equivalent to having two real singlets, $S = (S_1 + i S_2)/\sqrt{2}$,
, the only modification is an extra factor of 2 in $\Omega_{\rm DM}$ in
eq.\eq{OmegaDM},
as well as a doubling of the invisible higgs width.
The predictions for $M$ and $\sigma_{\rm SI}$ remain very similar as in the real case of fig.\fig{2}.
Furthermore the cosmological solution with $M>m_h/2$ becomes marginally allowed.

\section{Conclusions}
We have shown in fig.\fig{2} the range of DM mass $M$ and of the DM/nucleon cross section $\sigma_{\rm SI}$ `favored' (at weak confidence level)
by the weak excess reported by the CDMS experiment.
This motivated us to look for DM models that predict both $M$ and $\sigma_{\rm SI}$ to be within the CDMS range.
DM could be a scalar singlet with just a quartic coupling $\lambda$ to the Higgs.
Assuming that the free parameter $\lambda$ is determined from the cosmological freeze-out DM abundance,
$M$ and $\sigma_{\rm SI}$ are univocally predicted.
Such predictions depend on the higgs mass, not yet precisely known.
The predictions are shown in fig.\fig{2}.
Furthermore, the model predicts that higgs decays are mostly invisible, ${\rm BR}(h \to SS)\approx 0.8$.




\paragraph{Acknowledgements} 
We thank Gino Isidori for discussions.
This work is supported by the Swiss National Science Foundation under contract No. 200021-116372.


\small
\begin{thebibliography}{99}

\bibitem{CDMS} \hepart[0912.3592]{CDMS collaboration}.

 \bibitem{Xenon}
XENON collaboration,
  Phys.\ Rev.\ Lett.\  100 (2008) 021303
  [arXiv:0706.0039].


\bibitem{CDMSth} 
\hepart[0912.3797]{M. Kadastik, K. Kannike, A. Racioppi, M. Raidal}.  
\hepart[0912.4025]{A.~Bottino, F.~Donato, N.~Fornengo and S.~Scopel}. 
\hepart[0912.4221]{M.~ Ibe, T.~ T. Yanagida}.  
\hepart[0912.4511]{Q.~Cao, C.~ Chen, C.~ Li, H.~ Zhang}.  
\hepart[0912.4329]{R.~ Allahverdi, B.~ Dutta, Y.~ Santoso}.  
\hepart[0912.4510]{Q.~ Cao, I.~ Low, G.~ Shaughnessy}.  %
\hepart[0912.4507]{M.~ Holmes, B.~ Nelson}.  
\hepart[0912.4484]{M.~ Endo, S.~ Shirai, K.~ Yonekura}.  
\hepart[0912.4599]{K.~ Cheung, T.~ Yuan}.  
\hepart[0912.4701]{J.~  Hisano, K.~ Nakayama, M.~ Yamanaka}.  


\bibitem{DMS}
\hepart[0912.4722]{X. He, T. Li, X. Li, J. Tandean, H. Tsai}.


\bibitem{UllioCatena} \hepart[0907.0018]{R. Catena, P. Ullio}.

\bibitem{Lik} \art[0907.1891]{A. Ianni et al.}{Phys. Rev.}{D80}{043007}{2009}.

\bibitem{Zupan}
\hepart[0912.4264]{J.~ Kopp, T.~ Schwetz , J.~ Zupan}.  


   




\bibitem{McDonald} J. McDonald, Phys. Rev. D50 (1994) 3637.

\bibitem{Burgess}
\hepart[hep-ph/0011335]{C. Burgess, M. Pospelov, T. ter Veldhuis}

\bibitem{Barger}
\art[0706.4311]{V. Barger, P. Langacker, M. McCaskey, M. Ramsey-Musolf, G. Shaughnessy}{Phys. Rev.}{D77}{035005}{2008}.





 \bibitem{cosmoDM}
\hepart[astro-ph/0603449]{WMAP collaboration}.



\bibitem{EWPT} For the latest results, see the
LEP electroweak working group web page,
\url{http://lepewwg.web.cern.ch/LEPEWWG}.

\bibitem{hinv} \hepart[hep-ex/0107032]{LEP higgs working group}.

\bibitem{Ellis} \hepart[0801.3656]{J. Ellis, K.A. Olive, C. Savage}.

\bibitem{lattice} \hepart[0907.4177]{J. Giedt, A.W. Thomas, R.D. Young}.


\bibitem{MDM}
  \art[hep-ph/0512090]{M. Cirelli, N. Fornengo, A. Strumia}{Nucl. Phys.}{B753}{178}{2006}.



\end{thebibliography}
\end{document}